\documentclass[12pt,aaspp4, flushrt]{aastex}
\usepackage[table]{xcolor}
\usepackage[normalem]{ulem}
\usepackage{booktabs}
\usepackage{longtable}
\usepackage{amsmath}
\usepackage{footmisc}
\usepackage{lineno}
\usepackage{rotating}

\newcolumntype{P}[1]{>{\centering\arraybackslash}p{#1}}

{\par\begin{tabular}{P{0.5in}P{0.5in}P{0.85in}P{0.65in}}
  \hline
  \rowcolor[gray]{1} \\
  \textbf{Bandpass$^{a}$} & \textbf{MJD$^{b}$} & \textbf{Count rate$^{c}$} & \textbf{Uncertainty$^{d}$} \\ \\ \hline \\}
{\end{tabular}\par}

\begin{document}
\doublespace

\title{Optical/UV--to--X-ray Echoes from the Tidal Disruption Flare ASASSN-14li}
\author{\normalfont{Dheeraj R. Pasham$^{1,\dagger,*}$, S. Bradley Cenko$^{2,3}$, Aleksander Sadowski$^{1,\dagger}$, James Guillochon$^{4,\dagger\dagger\dagger}$, Nicholas C. Stone$^{5,\dagger}$, Sjoert van Velzen$^{6,\dagger\dagger}$, John F. Cannizzo$^{2}$}}\altaffiltext{1}{Massachusetts Institute of Technology, Cambridge, MA 02139} \altaffiltext{2}{NASA's Goddard Space Flight Center, Greenbelt, MD 20771}\altaffiltext{3}{Joint Space-Science Institute, University of Maryland, College Park, MD 20742, USA}\altaffiltext{4}{Harvard-Smithsonian Center for Astrophysics, Cambridge, MA 02138}\altaffiltext{5}{Columbia University, New York, NY 10027}\altaffiltext{6}{The Johns Hopkins University, Baltimore, MD 21218}\altaffiltext{$\dagger$}{Einstein Fellow}\altaffiltext{$\dagger\dagger$}{Hubble Fellow}\altaffiltext{$\dagger\dagger\dagger$}{Harvard ITC Fellow}\altaffiltext{*}{Corresponding author}

\begin{abstract}
We carried out the first multi-wavelength (optical/UV and X-ray) photometric 
reverberation mapping of a tidal disruption flare (TDF) ASASSN-14li. We find 
that its X-ray variations are correlated with, and lag the optical/UV 
fluctuations by 32$\pm$4 days. Based on the direction and the magnitude of 
the X-ray time lag, we rule out X-ray reprocessing and direct emission from a 
standard circular thin disk as the dominant source of its optical/UV emission.
The lag magnitude also rules out an AGN disk-driven instability as the origin of 
ASASSN-14li and thus strongly supports the tidal disruption picture for this 
event and similar objects. We suggest that the majority of the optical/UV 
emission likely originates from debris stream self-interactions. Perturbations 
at the self-interaction sites produce optical/UV variability and travel down to 
the black hole where they modulate the X-rays. The time lag between the optical/UV
and the X-rays variations thus corresponds to the time taken by these fluctuations 
to travel from the self-interaction site to close to the black hole. We further 
discuss these time lags within the context of the three variants of the 
self-interaction model. High-cadence monitoring observations of future TDFs will 
be sensitive to detect these echoes and would allow us to establish the origin of 
optical/UV emission in TDFs in general.
\end{abstract}

\section{Introduction}
When a star comes sufficiently close to a massive black hole (10$^{4-8}$ 
$M_{\odot}$) such that tidal forces exceed its self-gravity, it will be torn 
apart to cause a stellar tidal disruption flare (TDF; Hills 1975; Rees 1988). 
These events not only have the potential to uncover a large population of hidden 
massive black holes but could also enable us to understand how jets are launched 
and disks are formed.

One of the recent controversies about TDFs is, where does the majority 
of the optical/UV emission originate from in TDFs? (Piran et al. 2015 (P15); 
Roth et al. 2016; Dai et al. 2015, etc). In some of the best-studied
events, the optical/UV blackbody radius is much larger than the tidal radius and 
the temperature remains roughly constant over much of the evolution
(e.g., Chornock et al. 2014; Holoien et al. 2016b). Because TDFs are 
basically accreting supermassive black holes the natural inclination is 
to expect the same mechanisms operating in AGN, viz., X-ray re-processing, 
thermal emission from the inner radii of the accretion disk, etc, to also 
operate in TDFs. Here, we use publicly available multi-wavelength (optical/UV and 
X-ray) data from a recent TDF ASASSN-14li to constrain the origin of 
its optical/UV emission.

ASASSN-14li was discovered as an optical transient by the All-Sky 
Automated Survey for SuperNovae (ASASSN; Shappee et al. 2014) on 2014 
November 11 (Holoien et al. 2016a, H16). Based on its spatial coincidence with its host galaxy's 
nucleus, its peak luminosity of roughly 10$^{44}$ erg s$^{-1}$, its 
blue optical spectrum with broad and transient H$\alpha$ and He 
emission lines, it has been categorized as an event caused by the 
tidal disruption of a star by a supermassive black hole (H16). We carried out a photometric 
reverberation analysis of ASASSN-14li to find that the variations in 
its X-ray bandpass are correlated with, and lag the optical/UV  
fluctuations by 32 days. We describe the data and our 
analysis in \S 2 \& 3, and discuss the implications of our result 
for the likely origin of the optical/UV emission from this source in \S 4.

\section{Optical/UV and X-ray Observations} 
The multi-wavelength data used in this paper was acquired from 
two different facilities: the {\it Swift} satellite (Gehrels et
al. 2004) and the LCOGT (Las Cumbres Observatory Global Telescope) 
network (Brown et al. 2013). {\it Swift}'s X-Ray Telescope (XRT; 
Burrows et al. 2005) provided the X-ray (0.3-10 keV) data, while 
its UV Optical Telescope (UVOT; Roming et al. 2005) provided the 
UV data in the {\rm UVW2}, {\rm UVM2}, and the {\rm UVW1} filters 
with centroid wavelengths of 1928, 2246, and 2600 \AA, 
respectively (Poole et al. 2008). The UVOT also facilitated 
optical data in the {\rm U}, {\rm B}, and the {\rm V} bands with 
centroid wavelengths of 3465, 4392, and 5468 \AA, respectively 
(Poole et al. 2008). However, because the B and the V-band optical 
observations were poorly sampled for the second half of the {\it 
Swift}'s monitoring campaign (see Fig. 2 of H16), 
we also used published LCOGT's {\rm g}-band optical data (centroid 
wavelength of 4770 \AA). We re-analyzed all of the X-ray and the 
UVOT data from {\it Swift}. Our reduction and analysis procedures 
are described in detail below.

\subsection{X-ray Data Analysis}
{\it Swift} started monitoring the tidal disruption flare ASASSN-14li 
roughly eight days after its discovery on MJD 56983.6 (H16). It observed the source for 1-3 ks roughly once every 
three days for the first $\approx$ 260 days of the campaign. However,
after this time the target became sun-constrained to {\it Swift} 
and the monitoring campaign suffered from longer data gaps. Therefore, 
we limited our analysis to the first 270 days of post-outburst data at 
all wavelengths.

After extracting the clean eventlists from the raw Level-1 XRT 
products, we accounted for pile-up (Miller et al. 2015) by extracting 
X-ray (0.3-10.0 keV) source events from an annular region (rather than 
a circular region) centered on the source's centroid. As recommended by 
the XRT data analysis guide$\footnote{http://www.swift.ac.uk/analysis/xrt/pileup.php}$,
we modeled the PSF in each observation to estimate an inner exclusion
radius. The inner exclusion radius of the source extraction region 
varied between 0-15$\arcsec$ while the outer radius was fixed at 
50$\arcsec$. The background count rates were extracted from annuli 
centered on the source with inner and outer radii of 70$\arcsec$ and 
250\arcsec, respectively. Thus we extracted a pile-up and 
background-corrected source count rate from each {\it Swift} 
observation. We also corrected for bad pixels following the procedure 
outlined by the {\it Swift} data analysis 
guide$\footnote{http://www.swift.ac.uk/analysis/xrt/lccorr.php}$. 
Because ASASSN-14li's X-ray energy spectrum remains 
roughly constant during the first 250 days of the outburst (see Fig. 
3 of Miller et al. 2015 and Table S5 of Brown et al. 2016), the count 
rate serves as a good indicator of its intrinsic flux. 

\subsection{UVOT Data Analysis} The Ultra-Violet Optical Telescope 
(UVOT; Roming et al. 2005) on-board \textit{Swift} began observing 
ASASSN-14li at 10:08 UT on 30 November 2014. We downloaded the data 
from the HEASARC archive\footnote{See
\url{http://heasarc.gsfc.nasa.gov/cgi-bin/W3Browse/swift.pl}.}. In 
each of the six broadband filters (UVW2, UWM2, UVW1, U, B, and V), we 
corrected the image astrometry and stacked frames on a per-visit basis.  

We performed photometry on the stacked images using a 3\arcsec\ aperture, 
and corrected to the standard UVOT photometric system (Poole et al. 2008;
Breeveld et al. 2011). To account for contamination from the host 
underlying a host galaxy, a particularly important contribution to the 
observed emission in the redder filters, we utilize the coincidence-loss 
corrected host flux estimates from Miller et al. (2015) (see also H16 
and van Velzen et al. 2016). All the optical/UV light curves are consistent 
with H16. The cross-correlation results below are independent of 
host-subtraction. 

\section{Cross-Correlation Analysis}
In order to search for and quantify any correlation between the X-ray and 
the optical/UV variations, we employed the interpolated cross-correlation
function (ICCF) methodology as described by Peterson et al. (2004). We first
de-trended the X-ray light curve with a bending power-law model (red-curve 
in the top-left panel of Fig. 1). Each of the optical/UV light curves were
then de-trended with a power-law decay model. A sample fit for the UVW1 data 
is shown as a dashed red curve in the top-middle panel of Fig. 1. The X-ray 
residual was then cross-correlated with each of the seven optical/UV 
residuals. A sample ICCF between the X-ray and the UVW1 data is shown in the 
top-right panel of Fig. 1, while the rest of the ICCFs are shown in Figs. 2A 
\& 2B. It is evident that the CCFs peak between 25-35 days indicating that 
the X-ray variations lag the optical/UV fluctuations by 25-35 days. 

In order to confirm these CCFs, we also visually inspected them and compared 
the variability features in the X-ray data with the features in the optical/UV 
residual light curves. These are shown in the bottom-right panel of Fig. 1 and 
the right panels of Figs. 2A \& 2B. In each of these panels, common features in 
both the light curves are evident. Some of these CCFs also show evidence for an 
anti-correlation at zero lag. However, we do not consider it to be real as 
visual inspection of the light curve residuals did not reveal a strong evidence 
for an anti-correlation.

The X-ray and the lead-corrected optical/UV residual light curves in Figs. 1-2 
suggest that the correlation functions are driven by three X-ray variability 
features: two broad peaks (one around day 50 and the second around day 110) and 
one trough around day 80 in the X-ray data (see the top-right panel of Fig. 2A). 
Correspondingly, these features are also present in the optical/UV residual 
light curves but offset by $\approx$ -32 days (see the top-right panel of Fig. 
2A).  

In order to estimate the statistical significance of the correlations we
estimated the chance probability that the X-ray light curve would produce these same 
variability features evident in a given optical/UV light curve. For this purpose,
we simulated 10$^{5}$ white noise light curves to mimic the X-ray residual 
light curve. We constructed these synthetic X-ray light curves by drawing from 
a Gaussian distribution with a mean of zero and a standard deviation equal to 
the standard deviation of the observed X-ray residual light curve. We then 
evaluated simulated ICCFs by cross-correlating each of these simulated X-ray 
residual light curves with each of the optical/UV residual light curves. This 
way--for each X-ray--optical/UV ICCF--we built a distribution of the 10$^{5}$ 
correlation values at each lag. Using these distributions we extracted the 95 
(blue) and the 99\% (magenta) confidence contours shown in the top-right panel 
of Fig. 1 (see also Fig. 2A \& 2B). We then estimated the uncertainty in the 
peak and the centroid of the ICCFs using the Random Subset Selection (RSS) and 
the Flux Randomization (FR) procedures as described by Peterson et al. (1998) \&  
(2004). The resulting cross-correlation centroid (CCCD) and peak (CCPD) 
distributions for the case of X-ray vs UVW1 ICCF is shown in the bottom-left 
and the bottom-middle panels of Fig. 1, respectively. 

Using the binomial distribution formula we also estimated the global statistical 
significance of the X-ray variations lagging the optical/UV fluctuations to be 
$>$4.4$\sigma$. This was estimated based on the fact that five out of seven 
CCFs have the correlation peaks detected at greater than the 95\% confidence (see,
for example, Pasham et al. 2015 and Tombesi et al. 2010 within the context of timing 
and spectroscopy, respectively).

We also evaluated all the X-ray--optical/UV CCFs using the discrete 
cross-correlation function (DCF) algorithm as described by Edelson \& 
Krolik (1988). We find that the resulting DCFs are consistent with the 
ICCFs shown in Figs. 1-2. 

Because a CCF is simply a convolution of the auto-correlation function 
(ACF) with a transfer function, a lag between the X-ray and the optical/UV 
light curve is real only if it arises from the transfer function and not 
from the X-ray or the optical/UV ACF. To rule out an ACF origin for the 
lag we extracted the X-ray and the optical/UV ACFs to find that they do 
not have any statistically significant lag features except at zero lag. 

\section{Discussion}
First, we demonstrate that the measured optical/UV--X-ray lag in ASASSN-14li 
implies that none of the three primary mechanisms$\footnote[2]{Other 
mechanisms similar to stellar-mass black hole binaries may also play a role 
(e.g., Gandhi et al. 2010, and references therein) but on timescales much 
shorter than observed here.}$ that can drive X-ray variations to correlate 
with the optical/UV changes in AGN (e.g., McHardy et al. 2016; Edelson et 
al. 2015) are at play here. 
\begin{itemize}
\item{In the first mechanism, X-rays from close to the black hole 
scatter off material in the outer regions of the accretion disk or 
any other surrounding medium, lose energy, and get reprocessed into lower 
energy optical/UV photons (Edelson et al. 2015). In this case, the X-ray 
variations would lead the optical changes by a few days to a few tens of 
days depending on the distance to the outer disk (Morgan et al. 
2010)/reprocessing medium.}

\item{ But if a significant fraction of the optical/UV 
emission originates directly from the accretion disk as a thermal black body, 
then two kinds of correlations are possible:}

\subitem{(1) If the majority of the 
X-rays are produced in a corona that is powered by Compton up-scattering 
of near-UV seed photons from the thermal inner disk (Reynolds \& Nowak 
2003), then the X-ray variations would lag the near-UV fluctuations by a 
fraction of a day to a few days depending on the size of the corona and 
the inner disk (which are determined by the black hole mass (Ar{\'e}valo et 
al. 2005) and accretion rate).}

\subitem{(2) Also, in this scenario, optical/UV 
fluctuations due to accretion rate perturbations could propagate inwards 
on a timescale corresponding to the local viscous timescale (Ar{\'e}valo et
al. 2008). Then the optical/UV emission leads the X-rays by a few tens to 
millions of days (Breedt et al. 2009) again depending on the black hole 
mass and the accretion rate (Shakura \& Sunyaev 1973; Fig. 3).}

\end{itemize}

It is evident from Figs. 1-2 that the optical/UV emission ``leads'' the 
X-rays by 32$\pm$4 days. This rules out X-ray reprocessing as the dominant 
source of the variable optical/UV emission. Furthermore, assuming a black 
hole mass of 10$^{6.5\pm0.6}$ $M_{\odot}$ as derived from its host galaxy's 
bulge luminosity and the bulge's stellar velocity dispersion (H16; van Velzen et al. 2016), the seed photon scenario (case (1) above) 
is also unlikely as it would result in UV lead times of only a few thousands 
to a few tens of thousands of seconds. Below we show that direct 
optical/UV emission from a standard thin disk can also be ruled out.

ASASSN-14li's observed peak bolometric luminosity of roughly 10$^{44}$ erg 
s$^{-1}$ (H16) implies an accretion rate of $\lesssim$ 
0.2$^{+0.7}_{-0.2}$ $\dot{M}_{\rm Edd}$ (where $\dot{M}_{\rm Edd}$ is 
the Eddington accretion rate). Such a sub-Eddington rate implies that 
if an accretion disk formed quickly after the disruption it can be 
described by a geometrically thin, optically thick disk model of Shakura
\& Sunyaev (1973). The observed lags between the optical, the UV and the 
X-ray radiation, if coming from the viscous propagation of accretion rate 
fluctuations in such a thin disk, require, for possible black hole and 
disk parameters of ASASSN-14li, that the optical/UV emission come from 
within a radius of 50 $R_{g}$ (top-left panel of Fig. 3; $R_{g}$ is the 
gravitational radius). If emitted from outside this radius, the viscous 
timescales, and therefore the lags, would be much longer than the observed 
lags. On the other hand, the thin disk around a 10$^{6.5\pm0.6}$ $M_\odot$ 
black hole of the appropriate luminosity emits most of its optical and UV 
radiation from outside a radius of 500 $R_{g}$ (top-right panel of Fig. 3). 
As a result, the expected lags due to viscous propagation of perturbation 
in a thin disk are orders of magnitude longer than the observed ones (bottom   
panel of Fig. 3). This discrepancy rules out the standard, thin, circular 
disk solution. More importantly, this argues strongly that ASASSN-14li is 
not an accretion disk-driven AGN flare.

The slim disk accretion disk model (Abramowicz et al. 1988), describing 
super-Eddington accretion flows, predicts thicker accretion flow that can 
produce shorter viscous times comparable with ASASSN-14li's lags. However, 
ASASSN-14li is sub-Eddington and in this limit the slim disk model reduces 
to the fiducial thin disk solution (Strubbe \& Quataert 2009). Another 
possibility is that a circular disk of a different nature, thick and very 
hot (e.g., Coughlin \& Begelman 2014), forms. Such a disk can form in 
principle even for sub-Eddington accretion rates. However, no general 
solution for such a mode exists, and therefore in this picture the lag 
magnitude cannot be used to constraint the flow parameters. Finally, a 
truncated disk model--where the optical/UV are produced by the 
Rayleigh-Jeans end of an X-ray multi-color blackbody emission--cannot 
explain the observed high optical/UV luminosities (H16; 
but see below). 

Many recent numerical studies of tidal disruption events (e.g., 
Shiokawa et al. 2015; Hayasaki et al. 2016; Guillochon et al. 2014, G14; 
Bonnerot et al. 2017, B17) have shown that the in-falling stellar debris 
stream will undergo self-interactions because of relativistic 
apsidal precession (see, for example, Fig. 2 of B17). 
While these studies differ in their predictions for the evolution of 
the debris stream following the first self-interaction, they all agree 
that the role of self-interactions/shocks is to facilitate the process 
of circularization by removal of angular momentum. For instance, 
Shiokawa et al. (2015) and P15 argue that following 
the first stream self-interaction at the apocenter of the most bound 
debris, material spreads and rushes towards the black hole on the free-fall 
timescale. On the other hand, B17 suggested a discrete 
variant of the self-interaction model where the stream undergoes successive 
self-interactions--without much spreading in each interaction--and can 
eventually settle into a small ($\sim$ a few tens of gravitational radii) 
accretion disk. Alternately, the debris can instead lose only a small fraction of 
its energy in each self-interaction and follow more bound elliptical orbits. 
This would ultimately lead to the closing of the gap between the apocenter 
and the black hole, forming an elliptical accretion disk (G14).

The lags of the observed magnitudes can originate in all the three 
models (see Fig. 4 for schematics). In the shock-at-apocenter model the 
collisions of streams are likely to result in perturbations of the, 
otherwise relatively uniform and thin, tidal stream of gas, and 
subsequently vary the fraction of the tidal debris getting close to the 
black hole and being trapped or directly accreted there (e.g., 
S{\c a}dowski \& Narayan 2015). Such an inner small-scale accretion 
flow will be hot and will be modulated at the rate at which tidal 
debris--already affected by the self-interaction at large radii--returns 
to very close orbits. Therefore, heating up a clump of gas in the 
dissipation/interaction region resulting in the optical/UV emission will 
be followed by modulation of the energetic X-ray radiation coming from a 
region close to the black hole. Furthermore, under such circumstances, 
X-rays variations are expected to lag the optical/UV fluctuations by the 
infall time from the stream--stream interaction region. This lag is 
roughly equal to the half the orbital time of the debris orbit. For 
example, the shortest expected lag in this model would correspond to half 
the orbital period of the most bound orbit. This is roughly 11 days for a 
$10^{6.5}$ $M_{\odot}$ black hole (using Eq. 4 of P15), and 
is in agreement with the lags identified in this work. It is also plausible
that fluctuations propagate on a thermal timescale which would be slower 
than the free-fall timescale.

Bonnerot et al. (2017) argue that in the absence of strong magnetic stresses 
successive self-interactions can lead to the formation of an inner accretion 
disk $\sim$ a few tens of Rg. Time lags can also manifest in this 
series-of-discrete-interactions model (see Fig. 4). Each stream self-interaction 
can produce the optical/UV emission similar to the shock-at-apocenter model. 
Furthermore, fluctuations from the last self-interaction--just before the flow 
joins the inner accretion disk--can travel down to the black hole on a viscous 
timescale and modulate the X-rays. The viscous timescale in a thin accretion 
disk at a radius of a few tens of Rg is comparable to the observed lag (top-left 
panel of Fig. 3), and thus suggests that a TDF disk may be confined to within a few
tens of Rg.

We can also show that the expected properties of elliptical disks are 
consistent with the observed lags by constructing a simplistic model 
following G14. Recent numerical works have shown that
the flow can remain elliptical for roughly ten orbits (B17). Smaller mass black 
holes can maintain the ellipticity for longer duration. In such a 
model, we assume the emission follows the same radial profile as in a 
circular thin disk (Shakura \& Sunyaev 1973) but the extent of the emitting 
region is limited by the size of the elliptical orbit of the most bound gas. 
Under such assumptions, the radius dominating emission at a given frequency 
can be approximated as,

\[ R_{\nu} = \frac{4\pi\int_{6R_{\rm g}}^{R_{\rm max}} R^2 \Big( {\rm max}\big[1,\frac{R}{2R_{\rm T}}\big]\Big)^{-1} F_{\rm BB}(\nu,T_{\rm eff}(R)){\rm d}R }{4\pi\int_{6R_{\rm g}}^{R_{\rm max}} R \Big( {\rm max}\big[1,\frac{R}{2R_{\rm T}}\big]\Big)^{-1} F_{\rm BB}(\nu,T_{\rm eff}(R)){\rm d}R}\]

\noindent where $R_{\rm T}$ is the tidal radius for $1M_{\odot}$ star and $R_{\rm max}$ 
is the apocenter radius for the most bound orbit. $F_{\rm BB}(\nu,T_{\rm eff}(R))$ 
is the black body emission at frequency $\nu$ and temperature $T_{\rm eff}(R)$.

To convert the radius at which emission takes place to the lags between 
given wavelengths and X-rays, we assume that the corresponding viscous 
timescale is given through, $t_{\rm viscous}={\cal V}^{-1} t_{\rm ff}$, where 
${\cal V}$ is an ad hoc viscosity parameter, and $t_{\rm ff}$ is the local 
free-fall time. In an elliptical disk this ad hoc viscosity parameter is much higher than in a 
circular disk because we allow for larger ``effective'' $\alpha$-viscosity and 
disk thickness.

In Fig. 3 (bottom panel) we plot the expected lags in the elliptical model as a function of 
wavelength obtained assuming $M_{\rm BH}$, black hole mass, =10$^{6.5}M_{\odot}$,
accretion rate=$0.2\dot M_{\rm Edd}$, and ${\cal V}=0.05$. The obtained relation 
fits the observed values of the lags very well. If the assumed value of ${\cal
  V}=0.05$ is correct, the gas takes $\sim 20$ orbits to spiral down from given 
radius to the black hole, i.e., the gas would accreted quite rapidly towards the 
black hole. However, the dynamics of elliptical disks forming in tidal disruption events, 
and in particular their dissipation and emission profiles, are currently very 
poorly understood, and thus cannot be used to estimate the flow parameters 
unambiguously.

In summary, we have presented a cross-correlation analysis that suggests that ASASSN-14li's 
X-ray variations lag the optical/UV fluctuations by 32$\pm$4 days. We propose 
that at least three models can explain these lags: (1) shocks-at-apocenter model, (2) an
elliptical accretion disk model, and (3) series-of-discrete-interactions model. The shocks-at-apocenter 
model is similar to the model proposed by P5 but, in addition, 
we propose that the optical and the UV emission originates from physically distinct sites
as opposed to the single photosphere scenario suggested by P15. 
ASASSN-14li's data does not allow us to unambiguously detect the lags between 
the optical and the UV bands, but high-cadence optical and UV monitoring 
observations of a TDF would allow us to easily test this hypothesis, and 
establish the origin of the optical/UV emission in other TDFs.


\begin{figure}[h]

\begin{center}
\vspace{-0.75cm}
\includegraphics[width=6.4in, height=4.1in, angle=0]{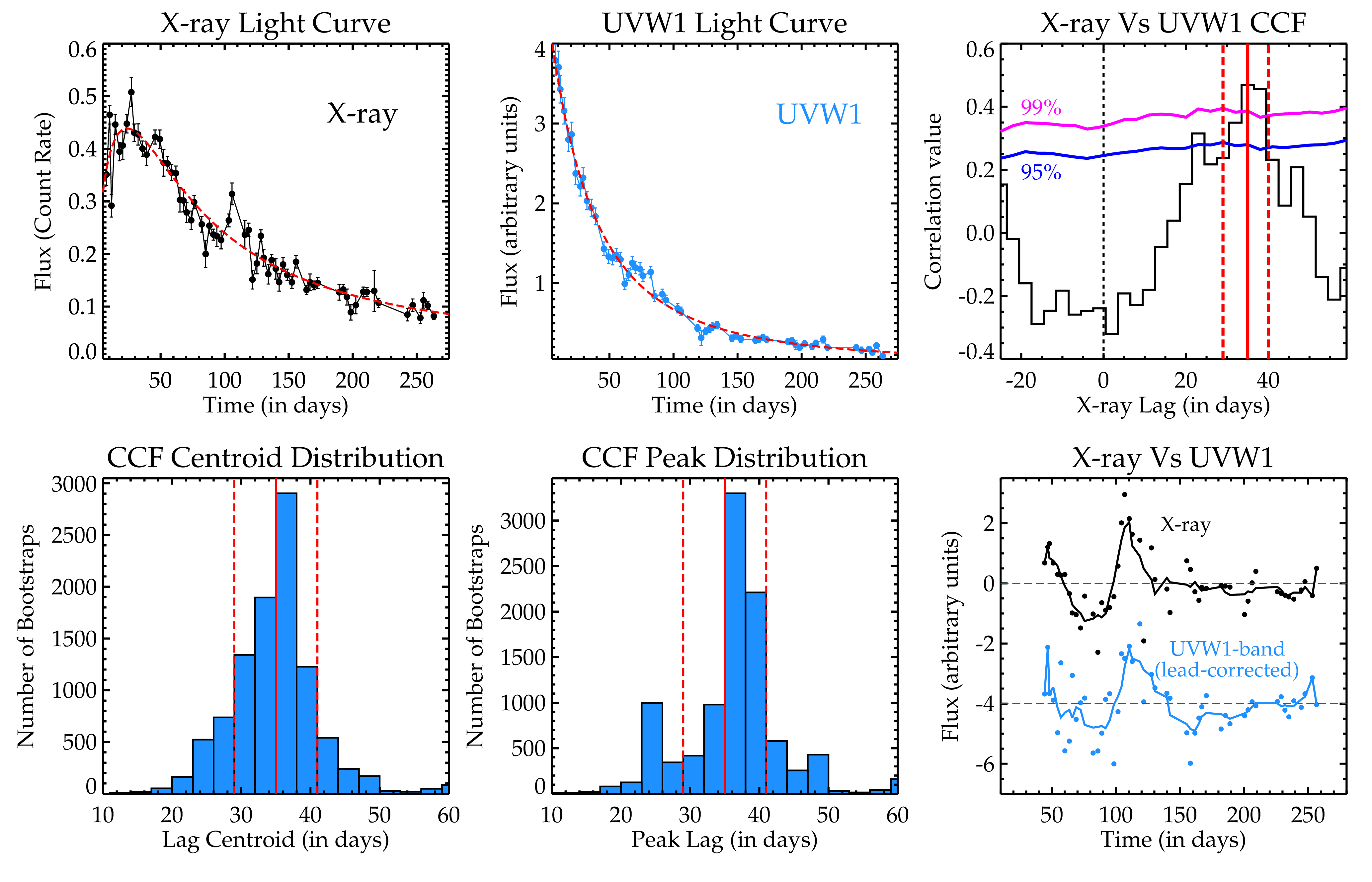}
\end{center}
\vspace{-.35cm} 
{\small {\bf Figure 1:} {\bf Top-Left:} ASASSN-14li's observed X-ray light curve (black 
data points) along with the best-fit bending power-law model (dashed red). {\bf 
Top-Middle:} ASASSN-14li's observed UVW1 light curve (blue) along with the 
best-fit decaying power-law model (dashed red). All the best-fit model parameters 
can be found in the machine-readable table MRT2.txt. {\bf Top-Right:} Interpolated 
Cross-Correlation Function (ICCF) between the de-trended X-ray and the UVW1 light 
curves. The 95 (blue) and the 99\% (magenta) white noise statistical confidence 
contours are also shown. The solid vertical red line is the median of the CCF's 
centroid distribution (35$^{+5}_{-6}$ days) shown in the bottom-left panel. The 
1$\sigma$ errorbars are shown as the dashed vertical red lines (same as in the 
bottom-left panel). {\bf Bottom-Middle:} ICCF's peak distribution. The solid 
vertical line is the median while the dashed lines are the 1$\sigma$ errorbars. 
{\bf Bottom-Right:} We compare the variability features in the X-ray and the 
UVW1 light curves. ASASSN-14li's UVW1 light curve is offset by 35 days (blue). 
The black data points show the x-ray light curve interpolated at the 
lead-corrected UVW1 epochs. Both the light curves were de-trended by subtracting 
a smooth function (top panels), leaving only the variability features. The solid 
curves are a running average of five neighboring points, to guide the eye. All the
X-ray and optical/UV photometric measurements can be found in the machine-readable 
table MRT1.txt.
}

\label{fig:figure1}
\end{figure}
\begin{figure}[h]

\begin{center}
\vspace{-0.75cm}
\includegraphics[width=4.2in, height=6.7in, angle=0]{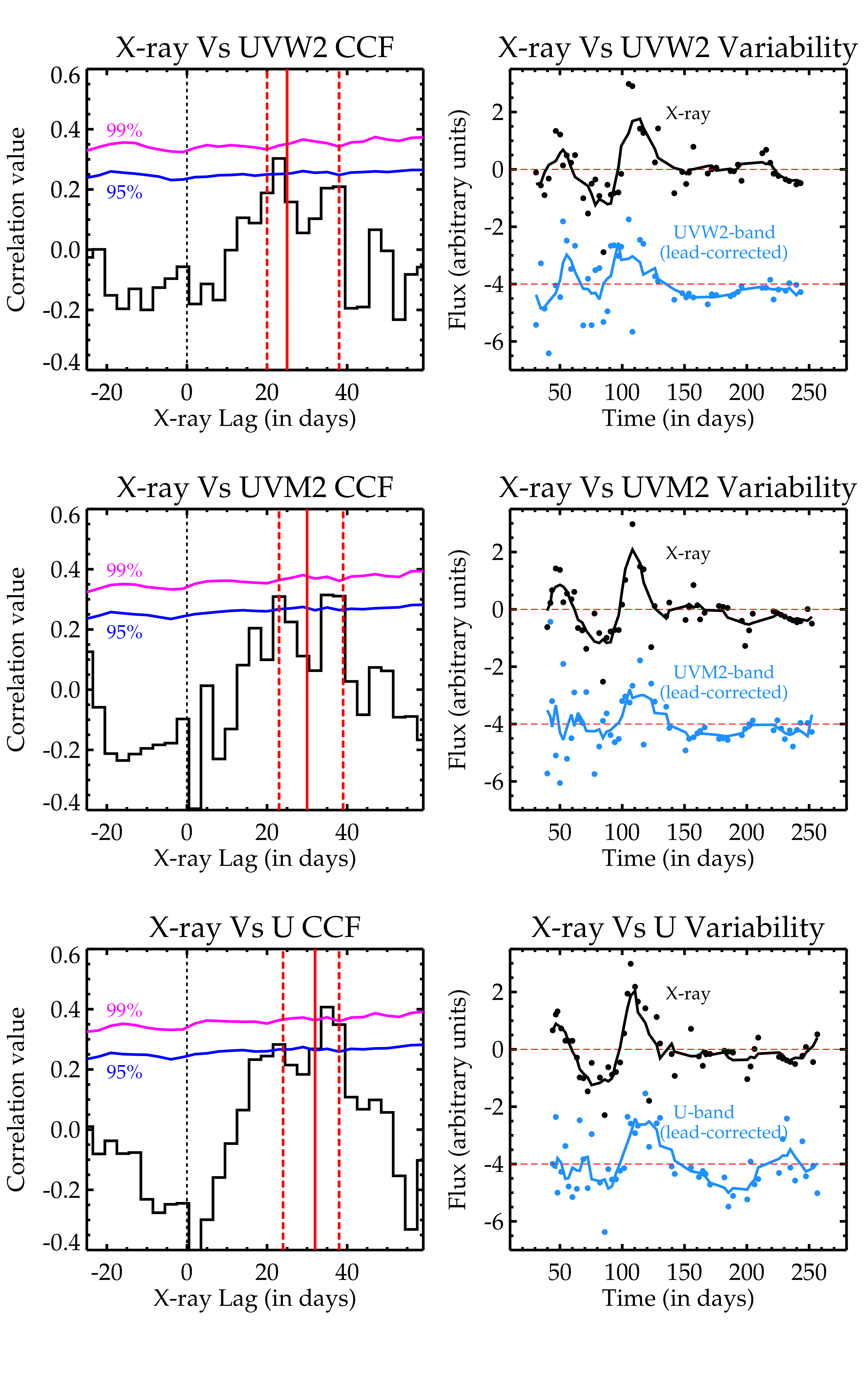}
\end{center}
\vspace{-1cm} 
\small{{\bf Figure 2A: } {\bf Left Panels:} X-ray vs optical/UV cross-correlation functions (CCFs). 
The two cross-correlated light curves are indicated at the top of each panel. The blue and the 
magenta curves are the 95 and the 99\% confidence contours, respectively. The solid vertical
red line is the median of the CCF's centroid distribution while the dashed vertical lines are
the 1$\sigma$ errorbars on the median. These values for the UVW2, UVM2, and the UVW1 ICCFs are 
26$^{+12}_{-6}$, 30$^{+8}_{-7}$, and 32$^{+6}_{-9}$ d, respectively. {\bf Right Panels:} We compare the variability features
in the X-ray light curve (black) with the optical/UV data (blue). The optical/UV light curves
were offset by their corresponding lead times. The solid curves are the running average of five 
neighboring points to guide the eye. Common features in both bands are evident in all the cases.
}

\label{fig:figure1}
\end{figure}
\begin{figure}[h]

\begin{center}
\includegraphics[width=4.2in, height=6.7in, angle=0]{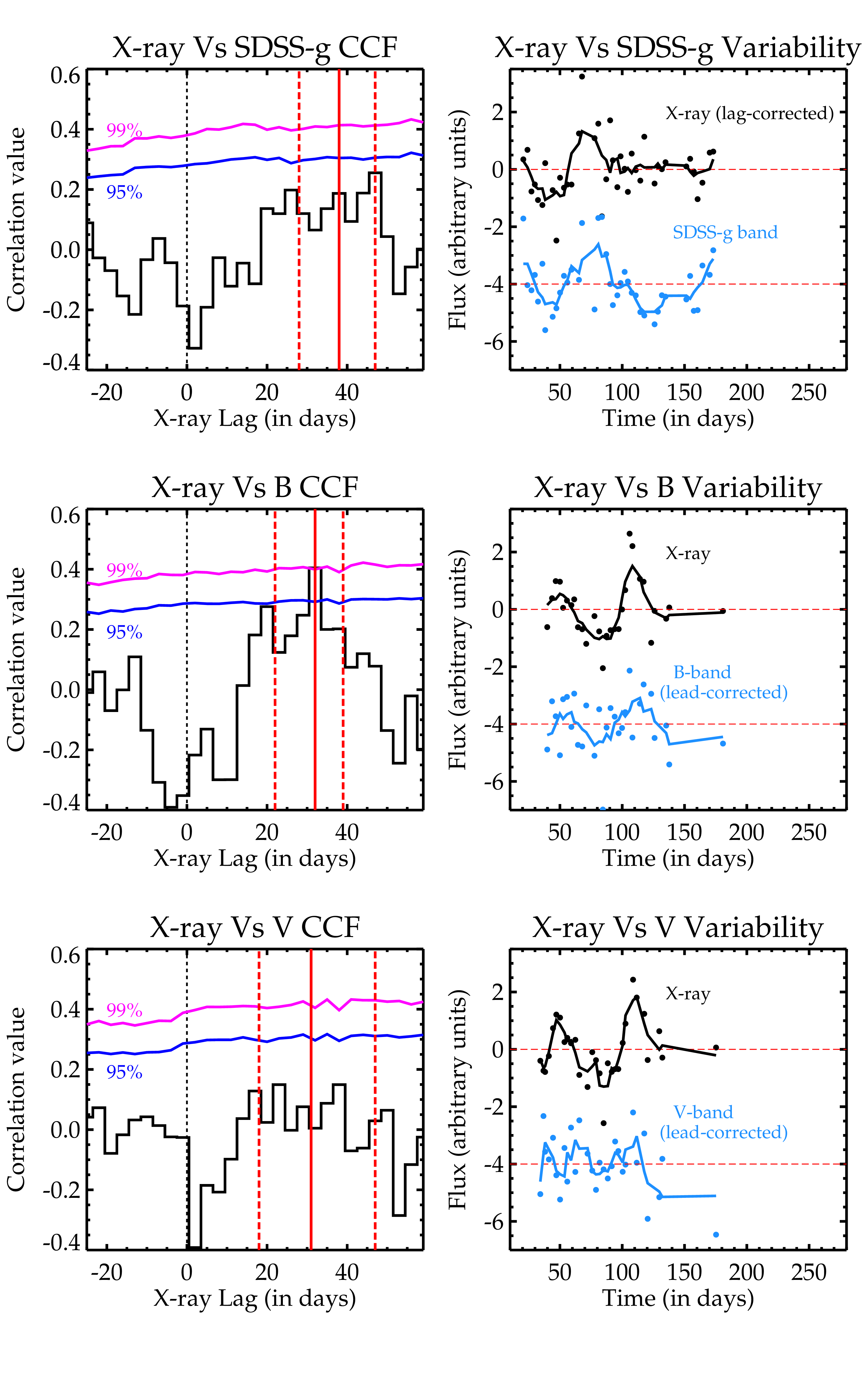}
\end{center}
{{\bf Figure 2B:} Same as Fig. 2A. The median of the CCF's centroid distribution for the SDSS-g, 
B, and the V band ICCFs are 37$^{+9}_{-10}$, 32$^{+7}_{-10}$, and 31$^{+16}_{-13}$ d, respectively. 
{\bf Top-Right:} Because SDSS-g band light curve was better sampled than the X-ray data, we offset 
the X-ray light curve by the lag implied from the CCF and interpolated the SDSS-g onto the X-ray epochs. 
}

\label{fig:figure3}
\end{figure}
\begin{figure}[h]

\begin{center}
\vspace{-0.5cm}
\includegraphics[width=6.5in, height=5.16in, angle=0]{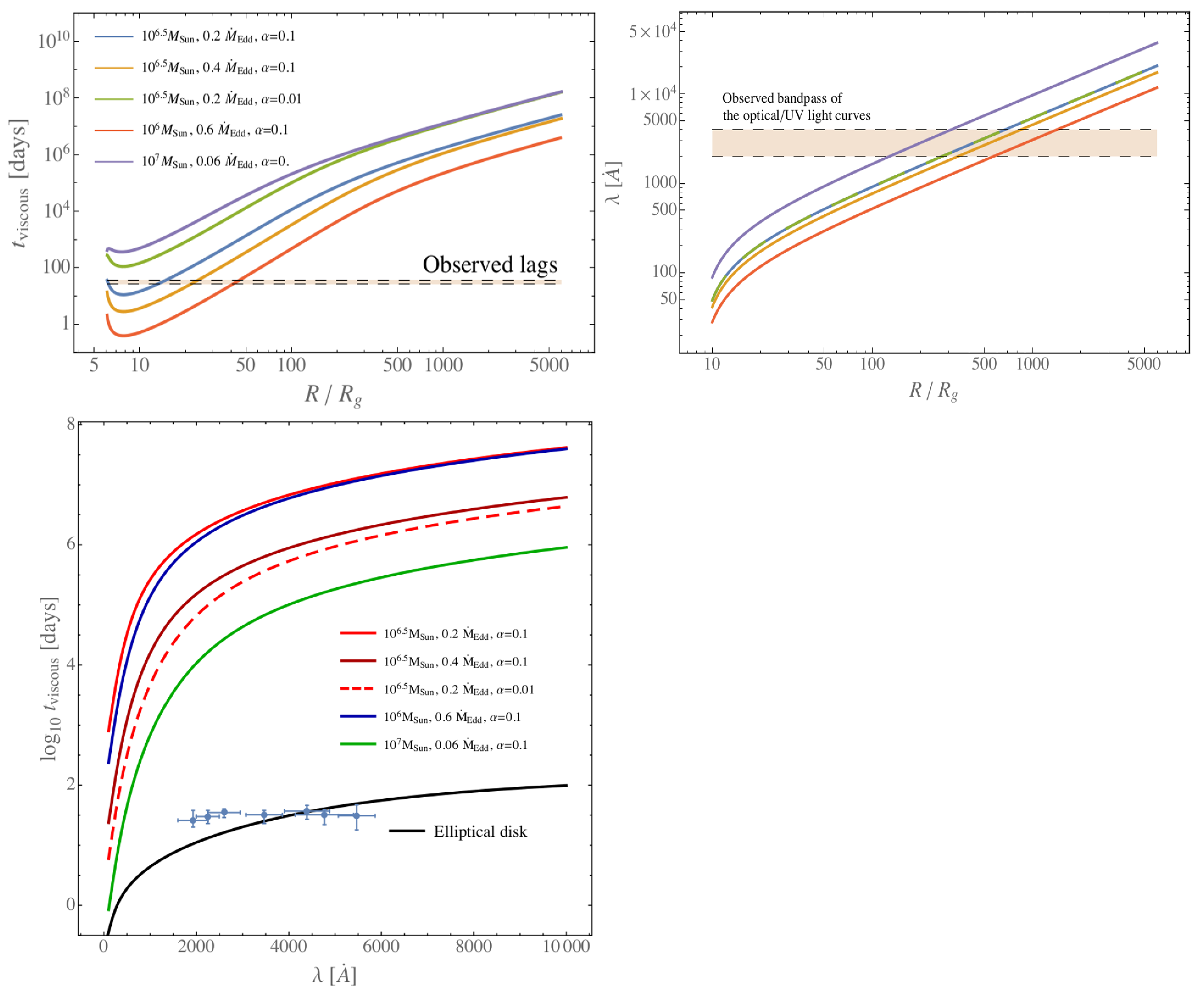}
\end{center}
\vspace{-0.5cm} 
{\normalsize{{\bf Figure 3:} {\bf Top:} The dependence of viscous timescale (top-left) and the 
wavelength of emission (top-right) on the radial distance in a Shakura \& Sunyaev (1973) thin 
disk are shown. The observed lags are inconsistent with a circular thin disk. The solutions for 
different black hole mass, Eddington ratio, and $\alpha$-viscosity parameters are shown. 
{\bf Bottom:} Viscous timescale vs wavelength of emission for a standard thin disk. The observed 
time lags, shown as blue data points, are orders of magnitude faster than expected from a circular
thin disk. A simplistic elliptical accretion disk model (G14) can reproduce 
the observed lags and is shown as a black curve. 
}}
\label{fig:figure4}
\end{figure}
\begin{figure}[h]

\begin{center}
\hspace{-0.5cm}
\includegraphics[width=6.65in, height=5.5in, angle=0]{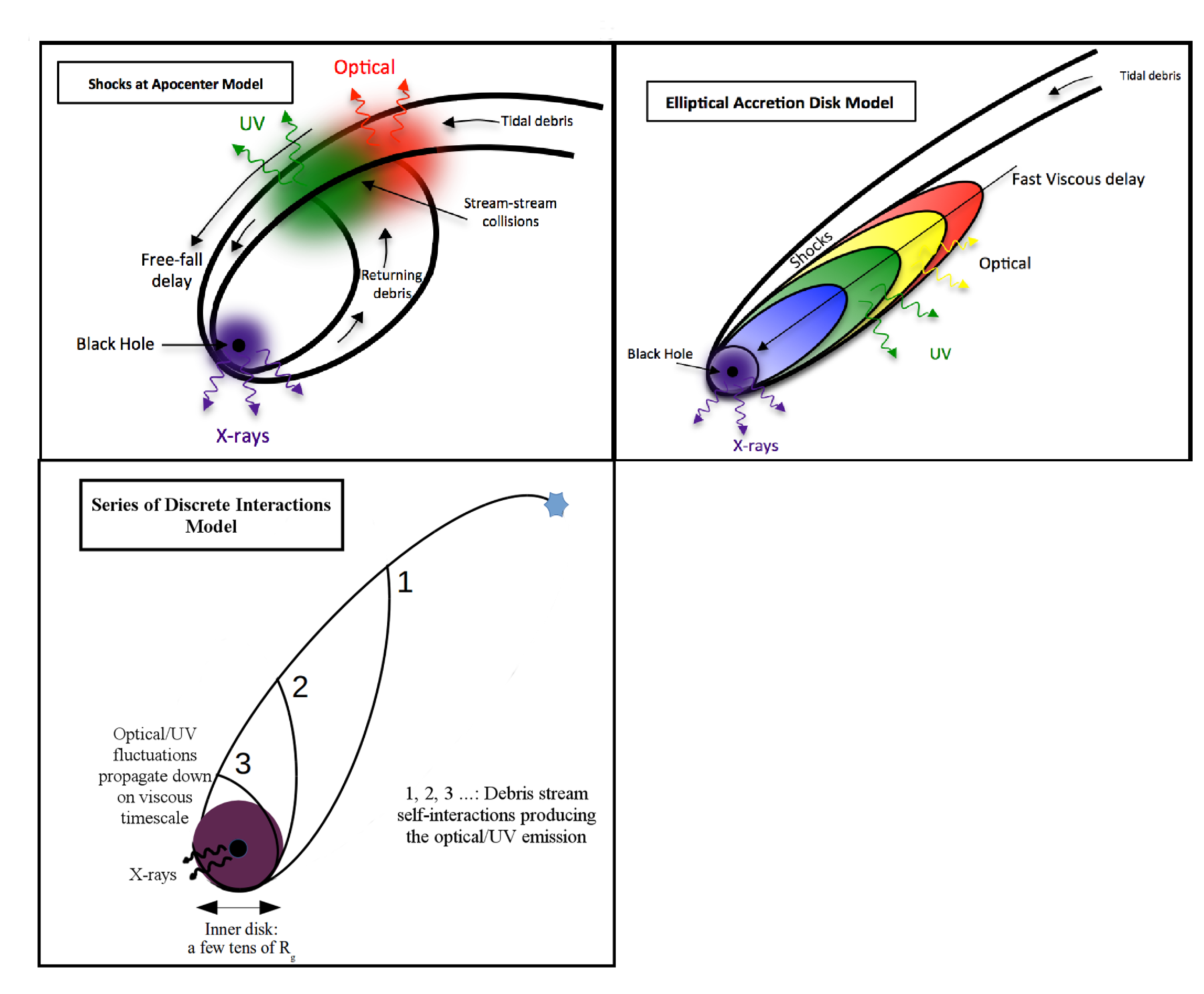}
\end{center}
{\normalsize{{\bf Figure 4:} Schematic of the three likely stream self-interaction 
scenarios for the origin of the optical/UV emission from ASASSN-14li. In the shocks-at-apocenter model 
(top-left), as clumps of debris trace the path shown, they first emit optical 
radiation, followed by UV and eventually--when near the black hole--modulate the 
X-rays. The time lag between the emission from these wavebands is dictated by the 
infall time between the optical, the UV and the X-ray sites. In the elliptical disk 
model (top-right), the time lags correspond to the local viscous timescale in the 
disk and are much faster compared to a circular disk of the same radial extent. In 
the series-of-successive-interactions model (bottom) fluctuations from the last 
self-interaction can propagate in the inner disk of a few tens of gravitational 
radii on the viscous timescale (see top-left of Fig. 3). 
}}
\label{fig:figure4}
\end{figure}

\end{document}